\documentclass[twocolumn,prb,showpacs,superscriptaddress]{revtex4-1}
\usepackage{amsmath}
\usepackage{amssymb}
\usepackage{bm}
\usepackage{graphicx}

\newcommand\pdag{{\phantom\dag}}

\begin{document}

\title{Exchange-interaction of two spin qubits mediated by a superconductor}
\author{Fabian Hassler}
\affiliation{JARA-Institute for Quantum Information, RWTH Aachen University, D-52056 Aachen, Germany}

\author{Gianluigi Catelani}
\affiliation{Peter Gr\"unberg Institut (PGI-2) and JARA-Institute for Quantum Information,
Forschungszentrum J\"ulich, D-52425 J\"ulich, Germany}

\author{Hendrik Bluhm}
\affiliation{JARA-Institute for Quantum Information, RWTH Aachen University, D-52056 Aachen, Germany}

\begin{abstract}
  Entangling two quantum bits by letting them interact is the crucial
  requirements for building a quantum processor. For qubits based on the spin
  of the electron, these two-qubit gates are typically performed by exchange
  interaction of the electrons captured in two nearby quantum dots. Since the
  exchange interaction relies on tunneling of the electrons, the range of
  interaction for conventional approaches is severely limited as the tunneling
  amplitude decays exponentially with the length of the tunneling barrier.
  Here, we present a novel approach to couple two spin qubits via a
  superconducting coupler. In essence, the superconducting coupler provides a
  tunneling barrier for the electrons which can be tuned with exquisite
  precision. We show that as a result exchange couplings over a distance  of
  several microns become realistic, thus enabling flexible designs of
  multi-qubit systems.
\end{abstract}


\maketitle

\section{Introduction}

Semiconductor based electron spin qubits have made significant progress
towards scalability. Single qubit gate fidelities demonstrated in some devices
meet the requirements for quantum error correction \cite{Morello, Dzurak},
with other approaches not being far behind \cite{Vandersypen, Tarucha,
Cerfontaine}. Two qubit gates have also been realized \cite{Shulman, Nowack},
and adequate fidelities seem within reach \cite{Dzurak}. However, all these
gates act over a very limited range of typically less than 1$\,\mu$m, which
severely constrains scalability as the resulting small qubit spacing leaves
little room for wiring and local control electronics.  For example, a surface
code architecture \cite{Mariantoni}, which is the currently most promising
main stream approach to error correction, requires a two-dimensional (2D)
lattice of qubits with nearest neighbor coupling.  One possible solution is to
use charge coupling.  Simulations indicate that the coupling range can be
extended to at least 10$\,\mu$m with floating electrostatic couplers
\cite{trifunovic} while maintaining a strength comparable to that demonstrated
in experiments with immediately adjacent qubits \cite{Shulman}. With this
coupling strength, the currently achievable level of charge noise \cite{Dial}
still leads to coherence times that are two orders of magnitude too short to
reach the required gate fidelities.  Another possibility is to transfer
electrons between qubits, e.g., using surface acoustic waves \cite{Hermelin,
MCNeil} or electrostatic gates \cite{Baart}. First evidence indicates that the
spin projection can be preserved during such a transfer \cite{Baart,
Bertrand}.  While one may hope that one can also achieve spin coherent
transfer, this remains to be shown experimentally. Furthermore, these
approaches entail a rather cumbersome complexity of the device and its
operation.

\begin{figure}[t]
  \centering
  \includegraphics{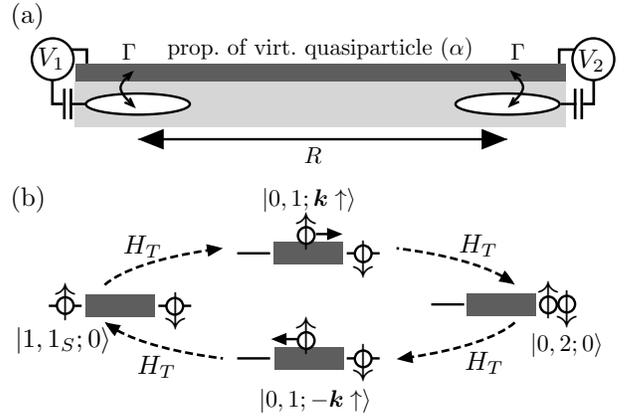}
  \caption{%
    (a) Sketch of the setup analyzed in the paper. Two semiconducting quantum
    dots (white ellipses) are tunnel-coupled with strength $\Gamma$ to a
    two-dimensional superconducting film (dark gray). The levels of the dot
    are tunable by the gate voltages $V_{1,2}$. (b) Steps in the virtual
    process leading to the exchange interaction. The initial conditions with
    one electron in each quantum dot is shown to the left. Given the fact that
    the electrons are in a singlet state, the following virtual process which
    is fourth-order in $H_T$ may take place: the electron in the left dot
    tunnels into the superconductor where it propagates as a quasiparticle
    above the gap. The electron enters the second dot which becomes doubly
    occupied. The two remaining tunneling processes reverse the path and take
    the electron back to the initial state. Note that that the reverse path
    could be also taking by the other electron which leads to a spin flip and
    explains the factor of two in Eq.~\eqref{eq:e_ex}.
  }\label{fig:setup}
\end{figure}

A very attractive remedy would be to directly extend the range of tunnel or
exchange coupling between localized electron spins representing a qubit. In
principle, this could be achieved with a very long and shallow tunnel barrier,
but in practice disorder would lead to localization on a scale of a few
$100\,$nm. Here, we theoretically analyze the possibility to use a
superconductor that is tunnel-coupled to the qubit electrons
(Fig.~\ref{fig:setup}) to mediate such coupling over extended distances
without being limited by localization.  Qualitatively, the main idea is to use
extended quasiparticle states for coupling while relying on the gap to freeze
out all the low-energy degrees of freedom in the coupler, thus suppressing
decoherence.

A key question is what coupling range and strength can be achieved with this
approach. To address this question for a simple model system, we compute the
exchange coupling between two distant electrons (e.g., localized in
semiconductor quantum dots) that are tunnel coupled to a 2D superconducting
ground plane and can be detuned electrostatically with respect to the latter.
We derive an expression relating the mediated coupling strength to that
achievable with direct coupling via the Green's function of the
superconductor, considering both the ballistic and disordered case. Using
realistic estimates of the relevant parameters, we find that a micron-scale
coupling with a practically useful strength of 10 to $100\,$MHz is achievable.

An important qualitative result is that the decay length of the coupling is
determined by the detuning between the quantum dot levels and the upper edge
of the superconducting gap. This detuning can be controlled with high precision via
gate voltages. Therefore, the decay length can be substantially larger than
the superconducting coherence length.  This result is in contrast to a recent
proposal \cite{leijnse:13} considering crossed Andreev Reflection as
coupling mechanism in a similar setting and finding a decay on the scale of
the coherence length.

The remainder of the paper is organized as follows. In Sec.~\ref{sec:setup},
we discuss the general setup and present its Hamiltonian.
Section~\ref{sec:clean} discusses the results for the case of a clean
superconductor. These results are contrasted in Sec.~\ref{sec:dis} with the
results for the disordered case. In Sec.~\ref{sec:estimate}, we present
realistic experimental parameters that allow an exchange coupling  over a
distance of several microns. Some  technical details about the disorder
average are moved to the Appendix.

\section{Setup}\label{sec:setup}

We want to study the exchange coupling strength in a setup where two
semiconducting spin qubits are coupled via a thin superconducting film, see
Fig.~\ref{fig:setup}(a).  The total Hamiltonian $H= H_D + H_\text{BCS} + H_T$
consists of three parts which we will discuss individually in the following.
The dots can be modeled by the  Hamiltonian $H_D = H_1 + H_2$ with
\begin{equation}\label{eq:ham_d}
  H_{j} = \epsilon_j n_{j}
  +\tfrac12 U_j  n_{j} (n_j -1).
\end{equation}
It involves the  operator $n_j= \sum_\sigma n_{j\sigma}$ counting the number
of electrons in dot $j$ where $n_{j\sigma} = d^\dag_{j\sigma}
d^\pdag_{j\sigma}$ and $d_{j\sigma}$ are fermionic operators.  The two dots
are tunnel-coupled via a Hamiltonian $H_T$ to a thin 2D superconducting film
of size $L\times L$. We measure the energies relative to the chemical
potential of the superconductor and   assume the tunability of the level
position $\epsilon_j$ by nearby gates $V_j$. The parameter $U_j>0$ describes
the Coulomb interaction due to the repulsion of multiple electrons on a single
dot.  We note that the eigenenergies $\varepsilon_{n_1,n_2}= \epsilon_1 n_1 +
\tfrac12 U_1 n_1 (n_1 -1) + \epsilon_2 n_2 + \tfrac12 U_2 n_2 (n_2-1) $ of the
dot Hamiltonian $H_D$ only depend on the occupation $n_1, n_2$ and not on the
spin state of the electrons. This is due to the absence of a magnetic field in
our description of the system. Note that, if needed, the application of a
(weak) magnetic field can be included perturbatively in the end. We seek a
situation where the states $|11\rangle$ at energy $\varepsilon_{1,1} =
\epsilon_1 + \epsilon_2$ and $|02\rangle$ at energy $\varepsilon_{0,2} =
2\epsilon_2 + U_2$ are almost degenerate with $\delta
\varepsilon=\varepsilon_{0,2} -\varepsilon_{1,1} >0 $ much smaller than the
typical energy spacing in the dots.  In this situation, we only have to take
the states $|1,1\rangle$ and $|0,2\rangle$ into account.  The near degeneracy
can be achieved by setting $\epsilon_1 = \epsilon_2 + U_2 - \delta
\varepsilon$.

We model the thin film of superconductor by the conventional BCS Hamiltonian
$H_\text{BCS} = \sum_{\bm k\sigma} E_k \beta^\dag_{\bm k\sigma}
\beta^\pdag_{\bm k\sigma}$.  The spectrum of the superconductor is given by
$E_k = ( \xi_k^2 + \Delta^2)^{1/2}$ with $\Delta>0$ the energy gap of the
superconductor and $\xi_k = \hbar^2 k^2/2m -\mu$; here, $m$ is the electron
mass and $\mu$ the chemical potential of the superconductor.  The fermionic
Bogoliubov operators $\beta_{\bm k\sigma}$ are related to conventional
electronic degrees of freedom via the unitary transformation
\begin{equation}\label{eq:bogoliubov}
  c_{\bm k\uparrow} = u_k \beta_{\bm k\uparrow} + v_k \beta^\dag_{-\bm k \downarrow},\quad
c^\dag_{-\bm k\downarrow} = -v_k \beta_{\bm k\uparrow} + u_k \beta^\dag_{-\bm k \downarrow}
\end{equation}
with the parameters $u_k,v_k\geq 0$ determined by $u_k^2 = 1-v_k^2 = \tfrac12
(1+ \xi_k/E_k)$. In the following, we denote with $|0\rangle$ the ground state
of the superconductor and correspondingly $|\bm k\sigma\rangle =
\beta^\dag_{\bm k\sigma} |0\rangle$ denote the (single-particle) excitations.

Coupling between the superconductor and the dot is provided by the tunneling
Hamiltonian
\begin{align}\label{eq:ht}
  H_T &= - t \sum_\sigma \Bigl[ c^\dag_\sigma(0) d^\pdag_{1\sigma} +
c^\dag_\sigma(R) d^\pdag_{2\sigma}  \Bigr] + \text{H.c.}, \nonumber\\
&= -\frac{t}{L} \sum_{\bm k\sigma}
\Bigl[ c_{\bm k\sigma}^\dag d^\pdag_{1\sigma}
+ e^{-i k_x R} c_{\bm k\sigma}^\dag d_{2\sigma}^\pdag \Bigr] + \text{H.c.},
\end{align}
where we have taken into account that the two dots are separated by a distance
$R$ (along the $x$-axis). For simplicity, we have assumed $t$ to be the same
in the two dots.  In the following, it will be useful to parameterize $t$ by
the tunneling rate $\Gamma/\hbar$ in the normal state with $\Gamma = 2\pi t^2
\rho_0$; here, $\rho_0 = m/2\pi \hbar^2$ denotes the density of states of the
normal state (per spin).

As we are considering the exchange effect mediated by the superconductor, an
important parameter will be the energy difference between the initial state
$\varepsilon_{1,1}= 2 \epsilon_2 + U_2 - \delta \varepsilon$ and the
intermediate state $\varepsilon_{0,1} + E_k = \epsilon_2 + E_k $ with the
electron in the superconducting wire.  For the latter to be an excited state,
we demand that $|\varepsilon_{1,1}- \varepsilon_{0,1}| < \Delta$.  In
particular, we are interested in a situation where $\varepsilon_{1,1}-
\varepsilon_{0,1} =\epsilon_2 + U_2 - \delta \varepsilon$ is smaller but not
much smaller than $\Delta$ (i.e., the level of dot 1 is tuned close to
the gap edge).  We parameterize this by the energy offset $M= \Delta -
(\epsilon_2 + U_2 - \delta \varepsilon)>0$ between the initial state with one
electron in each dot and the intermediate state where the electron of dot 1 is
transferred to the superconductor. Note that $M$ can be tuned independently of
$\delta \varepsilon$ by $\epsilon_2$.

In this situation, the dominant contribution for the exchange comes from a virtual
process where we start from $|1,1\rangle$ go over to a state $|0,1\rangle$
plus a low energy excitation in the superconductor, then we reach
$|0,2\rangle$ before we retrace the steps, see Fig.~\ref{fig:setup}(b). In
order that this exchange interaction can lead to a reduction of the ground
state energy, the spins of the electrons in the initial $|1,1\rangle$ state
have to be in a singlet as otherwise the $|0,2\rangle$ state is forbidden by
Pauli exclusion. The interaction thus assumes the form $H_\text{ex} = \tfrac14
J \bm \sigma_1\cdot \bm \sigma_2$ with $J>0$ the energy difference between the
singlet (which is lowered in energy) and the triplet state (which is
unchanged).

Assuming that the spins are in the singlet $|1,1_S\rangle = 2^{-1/2}
(d^\dag_{1\uparrow} d^\dag_{2 \downarrow}-d^\dag_{1\downarrow} d^\dag_{2
\uparrow} )|0\rangle$, we compute the lowering of the ground state energy in
fourth-order perturbation theory in the tunneling Hamiltonian $H_T$ (the
triplet energy does not change as discussed above).  We obtain $J=\alpha
\Gamma^2/\delta \varepsilon$ with the dimensionless coupling constant
\begin{align}\label{eq:e_ex} \alpha &=\frac1{\Gamma^2} \left| \sum_{\bm k
  \sigma} \frac{
    \langle 0,2 ; 0 | H_T |0,1; \bm k  \sigma \rangle \langle 0,1; \bm k
  \sigma| H_T | 1,1_S ;0\rangle}{\varepsilon_{1,1} -\varepsilon_{0,1} - E_k}
  \right|^2 \nonumber\\ &= \frac{1}{2\pi^2 \rho_0^2} \bigl|
  g(R;\Delta-M) \bigr|^2
\end{align}
where we have introduced the Green's function of the superconductor
\begin{align}\label{eq:g_sc}
  g(r; E) &= -i \int_0^\infty\!dt\,
  \langle 0 | c_{\sigma}^\pdag(\bm r)e^{i (E -H_\text{BCS}) t/\hbar}
  c^\dag_{\sigma}(0) | 0 \rangle \nonumber\\
  &=
  \int \frac{d^2 k}{(2\pi)^2} \frac{u_k^2 e^{i \bm k \cdot \bm
  r}}{E-E_k}
\end{align}
that we will compute in the following. Note that because we are interested in
values $E=\Delta-M< \Delta$, we do not need to distinguish between advanced
and retarded Green's functions. The exchange interaction is thus given by the
product of the bare result $J_0 = \Gamma^2/\delta \varepsilon$ which would be
achievable in the case the dots where in direct contact and a distance
dependent renormalization factor $\alpha<1$ describing the reduction due to
the finite spatial separation. Note that in Eqs.~\eqref{eq:e_ex} and
\eqref{eq:g_sc}, we have assumed that the superconductor is in the ground
state without and quasiparticle excitations present which requires that the
electron temperature is much smaller than the superconducting gap $\Delta$.

\section{Clean SC}\label{sec:clean}

In the case of a clean superconductor, it is straightforward to evaluate
\eqref{eq:g_sc}. Going over to polar coordinates and assuming $M \ll \Delta
\ll \mu$ and $k_F r \gg 1$ yields the semiclassical expression
\begin{align}\label{eq:g_sc_clean}
  g(r; E) &= \frac{\rho_0}{2} \int_{-\infty}^\infty\,
  d\xi_k \int_0^{2\pi} \!\frac{d\phi}{2\pi}\, \frac{
    e^{i  (k_F + \xi_k/\hbar v_F) r \cos \phi}}
    {E-\Delta-\xi_k^2/2 \Delta} \nonumber\\
    &= - \frac{\rho_0}{\sqrt{2\pi k_F r}} \mathop{\rm Re}
    \int_{-\infty}^\infty\!d\xi_k\,\frac{e^{i k_F r -i\pi/4+ i  \xi_k
      r /\hbar v_F
    }}
  {M+\xi_k^2/2 \Delta}
    \nonumber\\
  &= \rho_0 \left(\frac{\pi \Delta}
  {M k_F r} \right)^{1/2}\cos(k_F r +3 \pi/4) e^{-r/2 \xi}  ,
\end{align}
with the effective coherence length $\xi =\hbar v_F/ \sqrt{8 \Delta M}$ that
is a factor $(\Delta/M)^{1/2}\gg 1$ longer than the bare coherence length
$\xi_0 = \hbar v_F /\pi\Delta$.  In Eq.~\eqref{eq:g_sc_clean}, we have taken
into account that for the relevant part of the integral, we have $\xi_k
\approx \hbar v_F (k-k_F) \ll \Delta$ such that $u_k^2 \approx \tfrac12$ and
$E_k \approx \Delta + \xi_k^2/ 2 \Delta$.

Having evaluated the Green's function in the semiclassical limit, we are in
the position to evaluate the dimensionless coupling constant $\alpha = 2
\cos^2(k_F R + 3\pi/4) \alpha_0$ with
\begin{equation}\label{eq:alpha_clean}
  \alpha_0 =  \frac{\Delta} {4 \pi M k_F R}  e^{-R/\xi} .
\end{equation}
The $\cos^2$-dependence of $\alpha$ originates in our model from tunnel at
point (i.e., momentum-independent tunneling amplitudes). In a realistic
situation, the diameter $d$ of the dot is large such that $k_F d \gg 1$. In
this case, the result will be modified. In particular, the tunneling amplitude
will depend on the momentum mismatch between the dot and the superconductor. A
careful consideration of these effects turns out to be rather subtle and
beyond the scope of the present work, see, e.g.,
Ref.~\onlinecite{prada:04}. When comparing the results for a clean to a
disordered superconductor in Sec.~\ref{sec:estimate}, we use $\alpha_0$ to
ease comparison between the clean and the disordered case. This corresponds to
replacing $\cos^2$ by its typical value $1/2$.  In this way, we avoid the
dependence of the results on microscopic details that are relevant only in the
clean case.

\section{Disordered SC}\label{sec:dis}

In order to treat the case of a disordered superconductor it is useful to go
over from the Green's function $g(r;E)$ to the Gorkov Green's function
\begin{equation}\label{eq:gorkov}
  G(r; E) =  \int \!\frac{d^2k}{(2\pi)^2} \frac{ (E
    + \xi_k)  e^{i \bm k \cdot \bm
  r}}{E^2-E_k^2}
\end{equation}
as the latter has better analytical properties allowing to set up
perturbation theory in terms of Feynman diagrams.\cite{abrikosov} In the limit
$M\ll \Delta \ll \mu$ that we are interested in, we have
\begin{equation}\label{eq:ratio}
  \frac{E+\xi_k}{E^2-E_k^2} \approx \frac{u_k^2}{E-E_k}
\end{equation}
such that the two Green's functions $G$ and $g$ can be used interchangeably.
It is a well-known result \cite{abrikosov} that under an impurity average the
Gorkov Green's function is simply given by $\overline {G( r; E)} = G( r;E)
e^{-r/2\ell}$ with $\ell$ the mean free path in the disordered system.

In order to obtain the exchange coupling through a disordered superconductor,
we should average $\alpha$ over disorder. It is important to note that the
impurity average (denoted by the overline) cannot simply be performed
separately on the two Green's functions constituting $\alpha$. The reason is
that this neglects interference effects which are relevant for a disordered
sample with long phase-coherence length.  In fact, impurity scattering that
involves both Green's functions even becomes dominant at large distances due to
the emergence of a new length scale, which in the diagrammatic language is
subsumed by the ladder diagrams forming the diffuson.

Following ideas of Refs.~\onlinecite{smith:92,duhot:06}, we calculate the
diffuson approximation of  the product of Green's function entering $\alpha$
in the Appendix. We obtain the result ($E< \Delta$)
\begin{multline}\label{eq:diffuson}
  \int \!
  \frac{d^2 k}{(2\pi)^2} \overline{G(\bm k; E) G(\bm k-\bm q;E)}
  \approx \\ \frac{\pi \rho_0}{2} \frac{\Delta^2}{(\Delta^2-E^2) [
    (\Delta^2 - E^2)^{1/2} +\hbar
  D q^2/2]}
\end{multline}
with the diffusion constant $D= v_F \ell/2$; here, $G(\bm k; E)$ denotes the
Fourier transform of $G(\bm r;E)$. The expression \eqref{eq:diffuson} is valid
for weak-disorder with $k_F \ell \gg 1$ and for $q \ll k_F$.  Using this
expression, we can obtain the result for the disorder-averaged exchange
coupling ($q,\phi$ are the polar coordinates of $\bm q$)
\begin{align}\label{eq:alpha_dis}
  \overline{\alpha} &=\frac{\Delta}{8\pi \rho_0 M}
  \int\!\frac{d^2q}{(2\pi)^2} \frac{e^{i q R \cos\phi}}{ \sqrt{2 \Delta M}+
  \hbar D
    q^2/2} \nonumber\\
    &= \frac{\Delta  K_0( R/\xi_D) } {2\pi  M k_F \ell }  \approx
  \frac{\Delta\xi_D^{1/2}}{2(2 \pi)^{1/2} M k_F \ell R^{1/2} }e^{-R/\xi_D}
\end{align}
with $\xi_D =  \sqrt{\ell \xi/2} $.

When comparing the ballistic result \eqref{eq:alpha_clean} to the diffusive
result \eqref{eq:alpha_dis} there are two major differences: (\emph{i}) the
exponential decay is controlled by different length scales $\xi$ versus
$\xi_D$. (\emph{ii}) the algebraic decay of the former is given by $R^{-1}$
whereas the latter decays more slowly with the power $R^{-1/2}$.\cite{feinberg:03}

\section{Estimate of the coupling strength}\label{sec:estimate}

\begin{figure}
  \centering
  \includegraphics{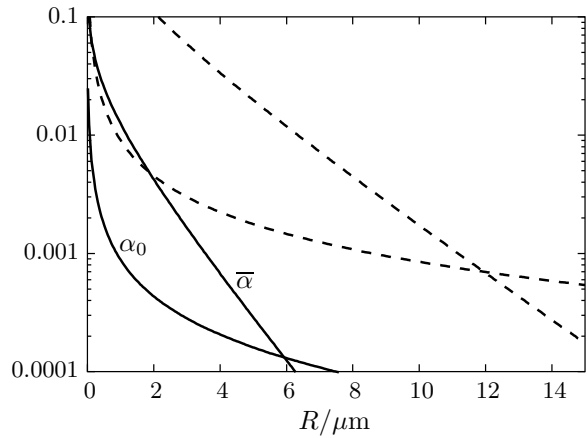}
  \caption{%
    Plot of the dimensionless coupling constant for clean aluminum ($\alpha_0$)
    and disordered aluminum ($\overline\alpha$) as a function of the distance
    $R$ of the spin qubits that are coupled. The plot shows both the results
    for $M/\Delta =5\times 10^{-3}$ (solid lines) and the results for
    $M/\Delta =5 \times 10^{-4}$ (dashed lines).
  }\label{fig:alpha}
\end{figure}

We would like to end by discussing realistic length scales $R$ over that the
superconducting film can be employed in order to exchange couple two spin
qubits. Starting from realistic values $J_0 \simeq \hbar (10 -100)\,$GHz for
the direct exchange coupling of two qubits, we aim at achieving a
dimensionless coupling constant $\alpha \simeq 10^{-3}$ in order to end up
with useful exchange coupling constants $J \simeq \hbar (10 -100)\,$MHz. For
the superconductor we propose  aluminum with a gap parameter $\Delta/k_B =
2.2\,$K which corresponds to a coherence length $\xi_0 = 2.3\,\mu$m for a
clean sample. Aluminum has a Fermi velocity $v_F =
2.0\times10^{6}\,\text{m}/\text{s}$ that implies a Fermi wave-vector
$k_F=17\,$nm$^{-1}$. The most crucial parameter which makes it possible to increase
$\alpha$ is the detuning $M$. Some of the requirements bounding $M$ from below
is that the detuning should be held stable over the time of exchange
interaction and that the smearing of the superconducting gap $\gamma$ (the
so-called Dynes parameter) should be much smaller than $M$. Recent experiments
have shown that $\gamma$ can be as small as $10^{-6} \Delta$.\cite{saira:12}
Given this input, we take a conservative choice of $M= 1\,\mu$eV which
corresponds to $M/\Delta = 5\times 10^{-3}$. As a result, we obtain the clean effective
coherence length $\xi = 36\,\mu$m and the dimensionless coupling constant
$\alpha_0$ of Eq.~(\ref{eq:alpha_clean}), see Fig.~\ref{fig:alpha},
\begin{equation}\label{eq:alpha_est}
  \alpha_0 = \frac{0.94\,\text{nm} }{R} e^{-R/36 \,\mu \text{m}},
\end{equation}
in a clean system as it is for example obtained by epitaxial
growth.\cite{chang:15} For a distance of $R=1\,\mu$m this evaluates to
$\alpha_0 = 9.1\times 10^{-4}$. We note that the value of $\alpha_0$ is
completely dominated by the prefactor. Thus, we expect that the analysis
becomes more favorable for the case of disordered aluminum as is obtained for
example by sputtering.

Extrapolating from Refs.~\onlinecite{mayadas:68,hiti:90,koshnick:07}, a
realistic value of the mean free path in aluminum is $\ell = 100\,$nm which
translates to a diffusive coherence-length $\xi_D = 1.3\,\mu$m.  This yields
\begin{equation}\label{eq:alpha_d_est}
  \overline\alpha = \frac{0.027\,\mu\text{m}^{1/2}}{R^{1/2}}
  e^{-R/1.3\,\mu\text{m}}
\end{equation}
and evaluates for $R=1\,\mu$m to the more favorable result $\overline\alpha =
1.3\times 10^{-2}$.

Figure~\ref{fig:alpha} shows that for distances $R$ smaller than a
characteristic distance $R^*$ the coupling via a diffusive superconductor is
larger than the one through a clean system \cite{notesmallR}. For $R>R^*$ this reverses.
For the typical case  $\ell \ll \xi$, the characteristic distance $R^*$ is
approximately given by
\begin{equation}\label{eq:rs}
  R^* \approx \frac{\xi_D}{2} \ln(\pi \xi/2 \ell).
\end{equation}
For $M/\Delta =5\times 10^{-3}$, we find numerically  that $R^*= 5.9\,\mu$m
and that $\overline\alpha$ is larger than $10^{-3}$ up to $R= 3.6\,\mu$m.

Provided accurate control over the gate voltages, the range of the exchange
coupling can be extended even further. For example, lowering $M$ by a factor
of 10, i.e., for $M = 0.1\,\mu$eV we obtain the dashed lines in
Fig.~\ref{fig:alpha}. We see that for distances smaller than $R^*=12\,\mu$m
the diffusive material leads to a stronger exchange interaction. Furthermore,
$\overline\alpha$ is larger than $10^{-3}$ for distances up to $R= 11\,\mu$m.

\section{Conclusion}

We have derived the strength of the exchange interaction between two spin
qubits which are connected via a 2D film of superconducting material acting as
a coupler. We have shown that the bare exchange interaction for two
neighboring dots is reduced by a dimensionless coupling constant $\alpha$
incorporating all effects of the finite distance $R$. We have presented
results for the case of  both  a clean and a disordered superconductor. We
have shown that for distances $R$ smaller than a characteristic distance
$R^*$, a diffusive superconductor outperforms the clean one due to the
prefactor in $\alpha$ for the former being a factor $(R/\ell)^{1/2}$ larger than for the latter.
We have shown that a diffusive superconductor with a moderate mean free path
of $\ell=100\,$nm enables to have useful exchange coupling strengths over a
distance of more thant $10\,\mu$m. In practice, it would be convenient to use
superconducting wires rather than an extended film to mediate the coupling.
In this case the further confinement would eliminate the $R^{-1/2}$ decay of
the prefactor in Eq.~\eqref{eq:alpha_dis}, leaving only the exponential  decay
and allowing substantially stronger coupling. This suggests that superconducting wires might be suitable for mediating an exchange coupling between quantum dots with a separation large enough to implement a 2D lattice. An additional advantage is that the coupling strength can be varied easily and rapidly by changing the electrostatic potential of the superconductor.

\begin{acknowledgments}
FH acknowledges financial support from the Alexander von Humboldt-Stiftung.
\end{acknowledgments}

\appendix

\section{Calculation of the Diffuson}

In this Appendix, we apply the ideas of Refs.~\onlinecite{smith:92,duhot:06}
to the specific case of disordered  2D superconducting film.  The calculation
of the impurity average has to be performed in Nambu space, because
the Green's function has  anomalous components  due to the superconducting
condensate. All the Green's function can be combined in the matrix Green's
function
\begin{equation}\label{eq:g_namb}
  \hat G(k; E) = \frac{  E + \xi_k \tau_3 + \Delta \tau_1}{E^2- E_k^2}
\end{equation}
acting via the Pauli matrices $\tau_i$ on the Nambu space. Note that the
connection to the Gorkov Green's function introduced in the main text is given
by $G(k;E) = \hat G(k,E)_{11}$. We assume throughout this section that
$0<E<\Delta$ such that the excitations are virtual and similar to the
Matsubara formalism we do not have to worry about retarded and advanced
Green's function. If required, the results for $E>\Delta$ can simply be
obtained by analytical continuation.

We are interested in performing an impurity average in the potential $\hat
V(\bm r) = v  \sum_i \delta^{(2)}(\bm r - \bm r_i) \tau_3$ where $\bm r_i$
denotes the position of the $i$-th impurity. The connection with the mean-free
path is given by the Born result $\hbar v_F/\ell = 2\pi n_i v^2 \rho_0$ where
$n_i$ is the density of impurities. The impurity averaged Green's function is
given by (see Ref.~\onlinecite{abrikosov})
\begin{equation}\label{eq:gav}
  \overline{\hat G (k; E)} = \frac{ \bar E + \xi_k \tau_3 + \bar\Delta \tau_1}
  {\bar E^2- \xi_k^2 - \bar \Delta^2}
\end{equation}
where $\bar E = \eta E$, $\bar \Delta =\eta \Delta$, and
\begin{equation}\label{eq:eta}
  \eta = 1 + \frac{\hbar v_F}{2 \ell (\Delta^2 - E^2)^{1/2}}.
\end{equation}

In order to calculate the disorder average of a product of Green's functions,
it is useful to introduce the four matrix diffusons ($\tau_0$ is the identity matrix)
\begin{equation}\label{eq:diffuson_mat}
  D_i(q) = n v^2
  \int \frac{d^2k}{(2\pi)^2} \tau_3 \overline{\hat G(\bm k;E) \tau_i
  \hat G(\bm k-\bm q; E)  }\tau_3.
\end{equation}
In terms of these, the combination of Green's functions in
\eqref{eq:diffuson} is
given by
\begin{equation}\label{eq:diffuson_eval}
  \int\! \frac{d^2 k}{(2\pi)^2} \overline{ G(\bm k; E) G(\bm k -\bm q;E)} =
  \frac{1}{2 n v^2} [D_0(q) + D_3(q)]_{11} .
\end{equation}

The dominant contribution to $D_i$ in the weak disorder limit originates from
ladder diagrams which keep the relative momentum $\bm q$ conserved. The zeroth
order term is given by
\begin{equation}\label{eq:d_0}
  D^{(0)}_i(q) = n v^2  \int\! \frac{d^2k}{(2\pi)^2} \tau_3 \overline{\hat G(\bm
  k;E)} \tau_i
  \overline{\hat G(\bm k-\bm q; E)  }\tau_3.
\end{equation}
As we are interested in small
relative momenta $q$, we can expand and obtain  $D^{(0)}_i(q) = a_i + b_i q^2 $
with
\begin{align}\label{eq:a_and_b}
  a_0&= n v^2 \rho_0 \int\!d\xi_k \,\tau_3  \overline{\hat G(k;E)}^2 \tau_3
  = \frac{\hbar v_F}{2 \ell}  \frac{\bar \Delta^2 - \bar\Delta \bar E \tau_1}
  {(\bar \Delta^2 -\bar E^2)^{3/2}}.
\end{align}
The expansion in $q$ is obtained by the replacement $\xi_{\bm k -\bm q} =
\xi_k - \hbar v_F q \cos \phi  $ with $\phi$ the angle between $\bm k$ and
$\bm q$. The first order in $q$ vanishes due to the integration over $\phi$. The
first non-vanishing contribution reads
\begin{equation}\label{eq:b}
  b_i= \frac{\hbar^2 v_F^2 a_i}{8 (\bar E^2 - \bar \Delta^2)}.
\end{equation}
As $D^{(0)}_0$ has a term proportional to $\tau_1$, we need to calculate
additionally $D_1^{(0)}$ in order to obtain a closed set of equations. We
obtain the expression
\begin{align}\label{eq:a1}
  a_1 =  \frac{\hbar v_F}{2 \ell}  \frac{\bar \Delta \bar E - \bar E^2 \tau_1}
  {(\bar \Delta^2 -\bar E^2)^{3/2}}.
\end{align}
For completeness, we give also the other expressions
\begin{align}\label{eq:a2_a3}
  a_2&= \frac{\hbar v_F}{2\ell} \frac{\tau_2}{(\bar \Delta^2 -\bar
    E^2)^{1/2}},
  &
  a_3&= 0.
\end{align}
For further convenience, we denote by $D^{(0)}_{i,j}$ the term in
$D^{(0)}_{i}$ proportional to $\tau_j$ such that  $D^{(0)}_i = \sum_j
D^{(0)}_{i,j} \tau_j$ by definition.

Summing the ladder diagrams is equivalent the Dyson type equations
\cite{Note1}
\begin{equation}\label{eq:dyson}
  D_i(q) = D^{(0)}_i(q)   +  \sum_j  D^{(0)}_{i,j} (q) D_j(q).
\end{equation}
The system is closed in the subspace $i,j \in \{0,1\}$. Solving the linear
system of equations, we obtain the result
\begin{equation}\label{eq:d0_full}
  D_i(q) =  \frac{D^{(0)}_i + (D^{(0)}_{0,1} D^{(0)}_{1,0} - D^{(0)}_{0,0}
  D^{(0)}_{1,1}) \tau_i}
  {1 - D^{(0)}_{0,0} - D^{(0)}_{1,1} - D^{(0)}_{0,1} D^{(0)}_{1,0}
  + D^{(0)}_{0,0} D^{(0)}_{1,1}}.
\end{equation}
Inserting the expressions for $D^{(0)}$ yields the final result
\begin{equation}\label{eq:d0_final}
  D_0(q) = \frac{\hbar v_F}{2\ell} \frac{\Delta^2 - \Delta E \tau_1}
  {(\Delta^2- E^2) [(\Delta^2
  -E^2)^{1/2} + \hbar D q^2/2]}
\end{equation}
valid for small $q$. For reference, we also give the result
\begin{equation}\label{eq:d1_final}
  D_1(q) =\frac{\hbar v_F}{2\ell} \frac{\Delta E - E^2 \tau_1}
  {(\Delta^2- E^2) [(\Delta^2
  -E^2)^{1/2} + \hbar D q^2/2]}.
\end{equation}
The diffuson $D_2$ only couples to itself, thus the solution of
\eqref{eq:dyson} is simply
\begin{equation}\label{eq:d2_final}
  D_2(q)= \frac{D_2^{(0)}}{1- D_{2,2}^{(0)}} = \frac{\hbar v_F}{2 \ell}
  \frac{\tau_2}{(\Delta^2-E^2)^{1/2} + \hbar D q^2/2}
\end{equation}
The $D_3$ diffuson vanishes such that only $D_0$ enters the expression
\eqref{eq:diffuson_eval}. From \eqref{eq:diffuson_eval}, we obtain the result
quoted in the main text.


\begin{thebibliography}{10}
\bibitem{notesmallR} We stress that these considerations are only valid in the
regime $R \gtrsim \xi_D$ that we are interested in.
\end{thebibliography}
\end{document}